\documentclass{egpubl}
\usepackage{egTechReport2019}

\ConferenceSubmission   

\usepackage{booktabs} 
\usepackage[T1]{fontenc}
\usepackage{dfadobe}  

\usepackage{cite}  
\BibtexOrBiblatex
\electronicVersion
\PrintedOrElectronic
\usepackage{egweblnk} 

\usepackage{algorithm}
\usepackage{algpseudocode}
\usepackage{adjustbox}
\usepackage{xspace}
\usepackage{relsize}

\usepackage[dvipsnames]{xcolor}
\usepackage{amsmath}
\usepackage{listings}
\usepackage{amssymb}
\usepackage{subcaption}
\usepackage{tabularx}
\usepackage{xcolor}
\usepackage{pgfplots}
\usepackage{tikz}
\usepackage{tabu}
\usepackage{mdframed}

\newmdenv[leftline=false,rightline=false,innertopmargin=+0.4cm]{topbot}

\definecolor{dblue}{rgb}{0,0,0.4}
\definecolor{bblue}{HTML}{4F81BD}
\definecolor{dred}{rgb}{0.4,0,0}
\definecolor{rred}{HTML}{C0504D}
\definecolor{dgreen}{rgb}{0,0.4,0}
\definecolor{ggreen}{HTML}{9BBB59}
\definecolor{dpurple}{rgb}{0.4,0,0.4}
\definecolor{ppurple}{HTML}{9F4C7C}
\definecolor{dyellow}{rgb}{0.4,0.4,0}
\definecolor{yyellow}{rgb}{1,1,0}
\definecolor{dlime}{rgb}{0.1,0.4,0}
\definecolor{llime}{rgb}{0.749,1,0}
\definecolor{dkingblue}{rgb}{0,0.1,0.6}
\definecolor{kkingblue}{rgb}{0,0.2,1}
\definecolor{dgray}{rgb}{0.3,0.3,0.5}
\definecolor{ggray}{rgb}{0.75,0.75,0.85}
\definecolor{dblack}{rgb}{0,0,0}
\definecolor{bblack}{rgb}{0.1,0.1,0.1}

\DeclareCaptionSubType*{figure}

\definecolor{dkgreen}{rgb}{0,0.6,0}
\definecolor{dkblue}{rgb}{0,0,0.6}
\definecolor{gray}{rgb}{0.5,0.5,0.5}
\definecolor{mauve}{rgb}{0.58,0,0.82}

\definecolor{commentgreen}{RGB}{2,112,10}
\definecolor{eminence}{RGB}{108,48,130}
\definecolor{weborange}{RGB}{255,165,0}
\definecolor{frenchplum}{RGB}{129,20,83}

\newcommand{\Test}[1]{\expandafter\hat#1}

\title[Adding Custom Intersectors to Visionaray]{
Adding Custom Intersectors to the C++ Ray Tracing Template Library Visionaray}

\author[Zellmann]{\parbox{\textwidth}{\centering
    Stefan Zellmann
    \thanks{zellmann@uni-koeln.de, Department of Computer Science, University of
    Cologne}\orcid{0000-0003-2880-9090}
    \quad
  }
}

\begin{document}

\teaser{ \centering \resizebox{0.98\textwidth}{!}{
\includegraphics[width=0.25\textwidth]{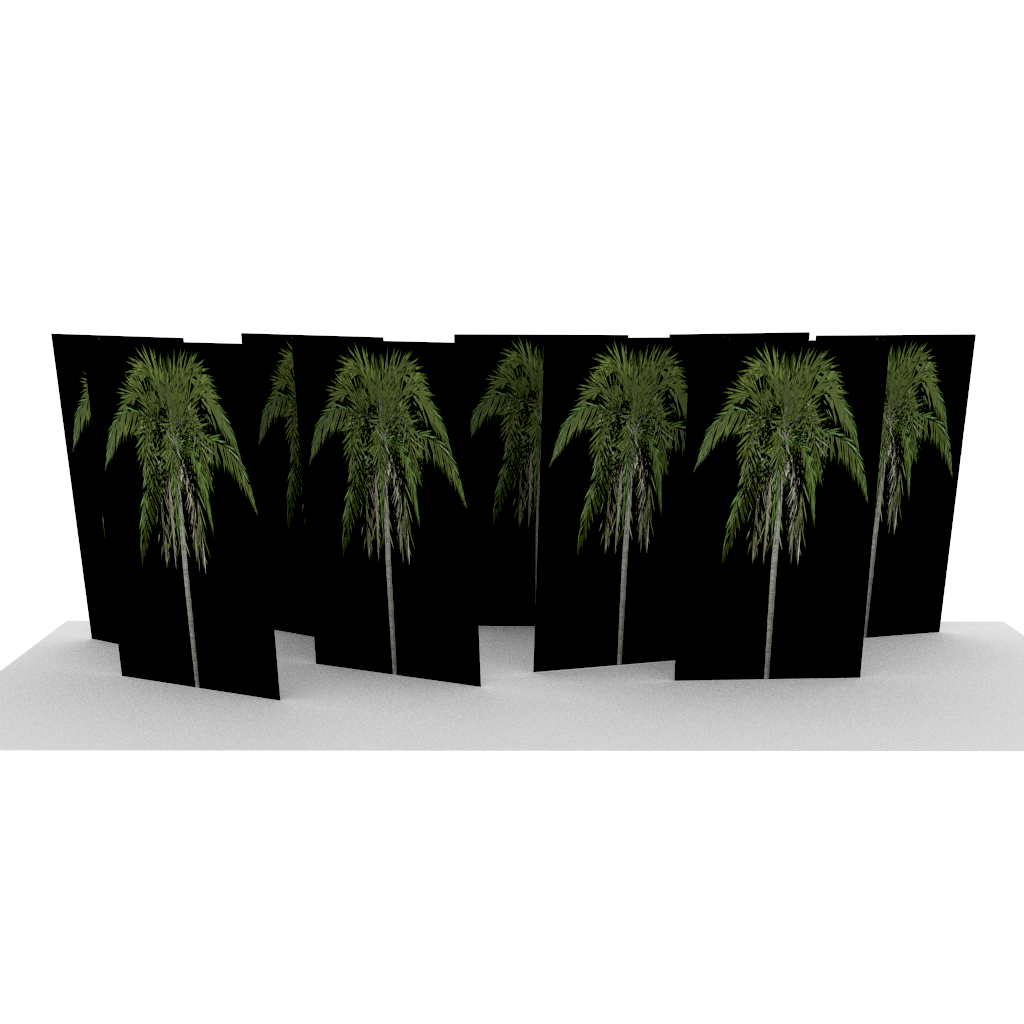}
\includegraphics[width=0.25\textwidth]{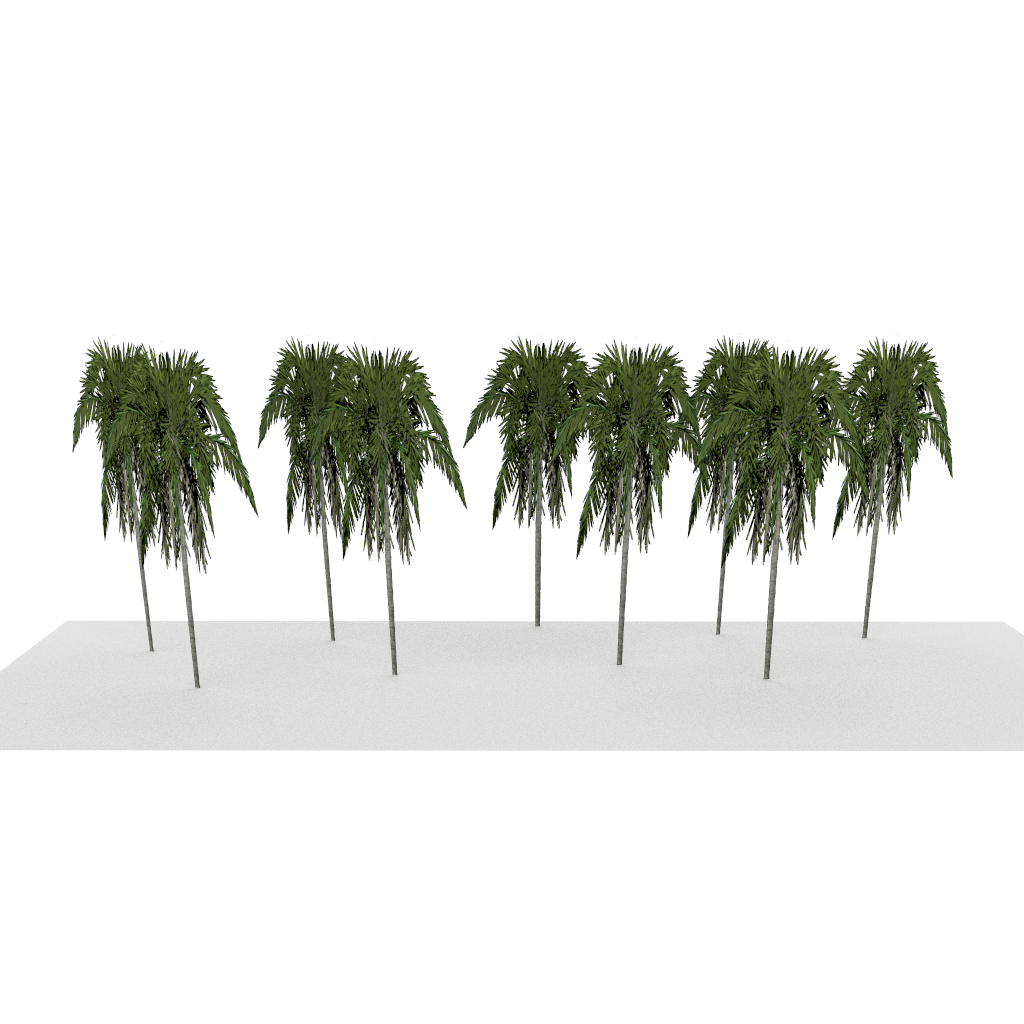}
\includegraphics[width=0.25\textwidth]{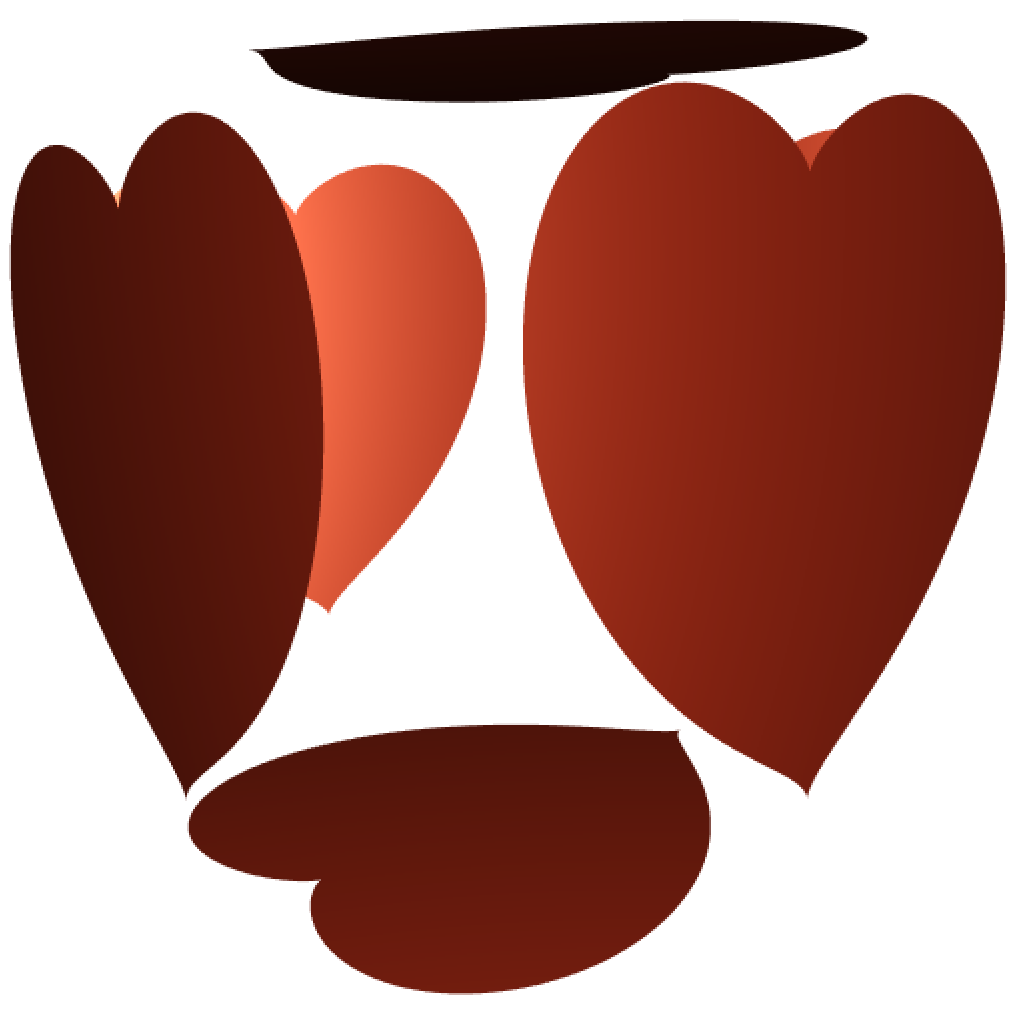}
\includegraphics[width=0.25\textwidth]{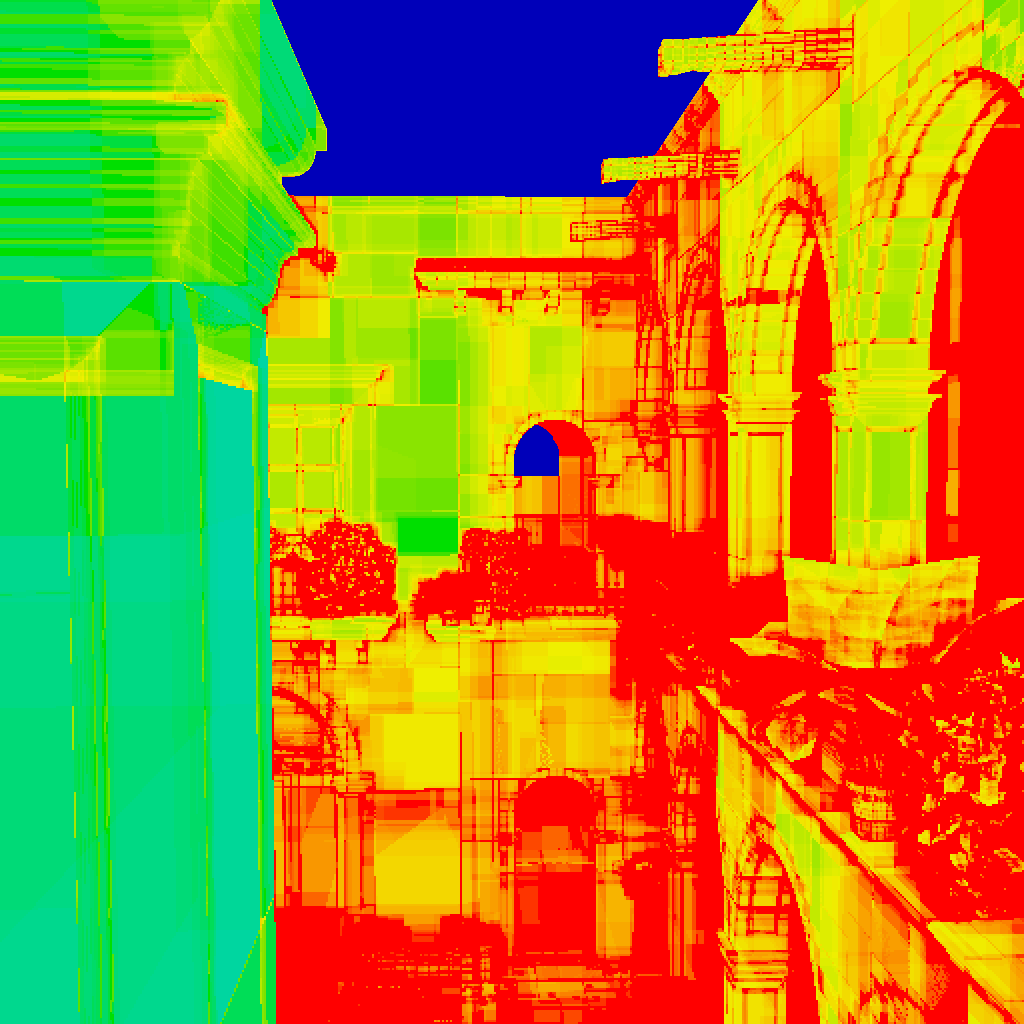}
 }
\caption{\label{fig:teaser}
Use cases where ray tracing algorithms were extended using custom intersectors.
From left to right: billboards, the alpha mask stored in the billboard images is
ignored. Second from left: the alpha mask from the billboard images is used to
conditionally continue BVH traversal when the surface has zero opacity near the
hit point. Second from right: procedural alpha mask applied using custom
intersectors. Right: debug image, the number of BVH nodes and the number of
primitives inside the encountered leaf nodes is used to generate a heat map.
This is done by intercepting the BVH traversal routine with a custom intersector
that counts the number of ray object interactions.
} }

\maketitle

\begin{abstract}
Most ray tracing libraries allow the user to provide custom functionality that
is executed when a potential ray surface interaction was encountered to
determine if the interaction was valid or traversal should be continued. This is
e.g.\ useful for alpha mask validation and allows the user to reuse existing ray
object intersection routines rather than reimplementing them. Augmenting ray
traversal with custom intersection logic requires some kind of callback
mechanism that injects user code into existing library routines. With template
libraries, this injection can happen statically since the user compiles the
binary code herself. We present an implementation of this ``custom intersector''
approach and its integration into the C++ ray tracing template library
Visionaray.
\end{abstract}

\section{Introduction}
Typical ray tracing libraries provide a default implementation for some
primitive type---e.g.\ for triangles---but allow the user to extend the library
in several ways. One such way might be support for completely new primitive
types that behave differently than the default primitive type. Often, the
functionality that the user desires however generally maps to the default
primitive type, but is instead an extension to the default behavior of that
type. One such example is support for alpha masks to selectively carve out areas
from planar surfaces based on a 2-d texture lookup. That functionality can be
implemented by extending the ray primitive intersection routine; the traversal
algorithm that tests the ray against a number of surfaces therefore first
performs the default intersection test, but instead of immediately reporting an
intersection first performs a lookup in the alpha texture and only reports the
hit if the lookup indicated the primitive was fully opaque at the
intersection position.

The functionality just described is usually implemented by extending the
intersection algorithm which will call some type of callback mechanism whenever
a potential hit was reported, and only report an actual hit if the callback
mechanism also reports an intersection. A default implementation might just
always report an intersection, no matter what the actual intersection position
was. This type of interface to the library can for example be found in Embree
\cite{wald:2014} where it is called an \emph{intersection filter} or in OptiX
\cite{parker:2010} where the functionality can be achieved by implementing a
custom \emph{any-hit program}.

In the context of libraries like Embree or OptiX that the programmer integrates
into her application by means of static or dynamic linking, this extension
mechanism must be evaluated at runtime: the library will perform some type of
runtime check if an intersection filter or any-hit program was registered and
only execute some custom functionality if a function pointer or some other means
to conditionally execute the user-supplied routine was properly initialized.

With template libraries like Visionaray~\cite{zellmann:2017b}, this check can
instead be performed at compile time and will incur zero cost if no custom code
was supplied by the user. As ray surface intersections are evaluated in the
innermost loop of the ray tracing algorithm, avoiding additional runtime checks
at this phase might improve overall performance.

In contrast to typical ray tracing libraries like Embree or OptiX, Visionaray is
a ray tracing template library where most of the functionality resides in C++
header files and is directly compiled into the user's application. This has the
advantage that the code can be inlined and optimized by the compiler in the
context of the application program. The approach also has certain disadvantages,
such as increased compile time for the application programmer, or the fact that
the application programer needs to make sure and rely on her compiler that the
program is properly optimized. An advantage of the approach however is that code
that is not needed is never actually compiled into the application and can thus
not have a negative impact for example on instruction cache utilization.

In this paper we describe the integration of \emph{custom intersectors} as an
application programming interface for static routines that the user implements
and passes to Visionaray at compile time.

\section{Related Work}
Visionaray provides support for several features that are required to develop
ray tracing algorithms, such as a streamlined texture interface that can be
leveraged on both CPUs with vector instructions as well as on NVIDIA GPUs
\cite{zellmann:2015}. Visionaray supports several types of intersection queries,
including the multi-hit query type~\cite{zellmann:2017}. Support for multiple
primitive, material or light types in the same ray tracing program is provided
by means of compile type polymorphism. The effectiveness of that approach was
thoroughly evaluated in~\cite{zellmann:2017c}. We used Visionaray and its
various library subsystems such as the vector math system or SIMD library
component to implement several algorithms, e.g.\ the ones from
\cite{zellmann:2018}, \cite{zellmann:2019},~\cite{zellmann:2019b}, and
\cite{zellmann:2019c}.

\section{Integration of Custom Intersectors into Visionaray}
In this section we first describe the application programming interface (API) by
which custom intersectors can be used by the application programmer, and then
provide details about how that was internally implemented in the library.

\subsection{Application Programming Interface}
Visionaray has a \emph{customization point} interface where the user can
overwrite or augment behavior by implementing free functions for custom types.
Custom geometric primitives e.g.\ can be added by implementing a set of free
functions, one of them being the \texttt{intersect} function that tests if a ray
intersects the primitive:
\begin{lstlisting}[
    language=C++,
    captionpos=b,
    label=lst:generic]
template <typename Ray>
hit_record<Ray, primitive<unsigned>> intersect(
        Ray const& ray,
        custom_primitive const& prim
        );
\end{lstlisting}
\texttt{custom\_primitive} in this case is a user-defined type, and the
\texttt{hit\_record} template contains a member variable \texttt{hit} that
indicates whether an intersection occurred or not.

\emph{Custom intersectors} allow to augment the behavior of \emph{existing}
primitive types like triangles; the user may e.g.\ be perfectly fine with the
triangle intersection routine as such (and may also wish to reuse the existing
builtin triangle type instead of implementing a completely new one), but wants
to add an alpha mask from an image texture.

The API for that consists of deriving from the \texttt{basic\_intersector}
template using the ``curiously recurring template pattern'' (CRTP):
\begin{lstlisting}[
    language=C++,
    captionpos=b,
    label=lst:generic]
struct custom_intersector
    : basic_intersector<custom_intersector> {
    using basic_intersector<
        custom_intersector
        >::operator();

    // Implementation goes here
    ...
\end{lstlisting}
The user then implements her own \texttt{operator()} as a member function, with
the signature of the \texttt{intersect} function from before, with the ray as
first and the custom primitive as second parameter:
\begin{lstlisting}[
    language=C++,
    captionpos=b,
    label=lst:generic]

    ...
    template <typename Ray>
    hit_record<Ray, primitive<unsigned>>
    operator()(
            Ray const& ray,
            custom_primitive const& prim
            )
    {
        auto hr = intersect(ray, prim);

        // use hit record, e.g. barycentrics
        // for texture lookups
        ...

        // manipulate the hit record
        hr.hit &= ...;

        // after manipulation, return
        return hr;
    }
};
\end{lstlisting}
Visionaray supports several visibility queries; usually, the ray is tested
against a \emph{bounding volume hierarchy} (BVH) built over some primitives, and
either the closest hit point (closest-hit query), the first encountered hit
point (any-hit query), or the first $N$ hit points (multi-hit query) are
returned. The interface for the various queries is similar and exemplarily is
presented here for the closest-hit query:
\begin{lstlisting}[
    language=C++,
    captionpos=b,
    label=lst:generic]
// Default overload
template <
    typename Ray,
    typename PrimIterator
    >
auto closest_hit(
        Ray          ray,
        PrimIterator first_prim,
        PrimIterator last_prim
        )
    -> decltype(intersect(ray, *first_prim));

// Overload w/ custom intersector
template <
    typename Ray,
    typename PrimIterator,
    typename Intersector
    >
auto closest_hit(
        Ray          ray,
        PrimIterator first_prim,
        PrimIterator last_prim,
        Intersector  intersector
        )
    -> decltype(intersect(ray, *first_prim));
\end{lstlisting}
Visionaray comes with a set of default kernels that implement various algorithms
like path tracing or plain primary visibility ray casting; those algorithms make
use of the aforementioned visibility queries and can be passed a custom
intersector using the scheduling parameters (for more details
see~\cite{zellmann:2017b}).

\subsection{Implementation Notes}
Visionaray's bounding volume implementation that is based on the
\texttt{while-while} traversal scheme from~\cite{aila:2009} calls
\texttt{intersect} twice. A high-level representation of the traversal scheme
looks like this:\\
\begin{topbot}
\begin{algorithmic}
\Procedure{intersect}{ray, BVH}
    \While {ray not terminated}
        \While {node is inner}
            \State \Call{intersect}{ray, node.bounds}
        \EndWhile

        \Comment{Found a leaf}
        \While {node contains untested primitives}
            \State HitRecord $\gets$ \Call{intersect}{ray, node.prims++}
        \EndWhile
    \EndWhile
\EndProcedure{}
\end{algorithmic}
\end{topbot}
The traversal function, at compile time, is passed the custom intersector class,
and the two calls to \texttt{intersect}---the one that tests the ray against the
bounding box of the BVH nodes and the one that tests against the individual
primitives---are statically replaced with the calls to \texttt{operator()}
provided by the custom intersector.

Also note how \texttt{intersect} inside the BVH traversal routine is not only
called for each geometric primitive, but also when the ray is tested against the
BVH nodes' bounds (which have type \texttt{aabb} with Visionaray). Custom
intersectors thus cannot only be used to intercept the behavior of ray vs.
primitive intersection, but also that of testing rays against BVH nodes. This
can be accomplished by adding an \texttt{operator()} overload to the custom
intersector that takes an \texttt{aabb} as second parameter.

An implementation detail worth mentioning is that the ray vs. BVH traversal
function in Visionaray is also called \texttt{intersect}. The reasoning behind
this is that the visibility queries (\texttt{closest\_hit}, \texttt{any\_hit}
and \texttt{multi\_hit}) will iterate linearly over a list, where BVHs may
themselves act as (compound) primitives. This allows us to easily implement
object instancing, where the BVH will store BVHs as primitives, and where the
object hierarchy may optionally have more than one root node. Conversely, in
certain cases the user might decide that a BVH is not required and just pass
iterators to a linear list of primitives to the query routines. Special care is
necessary to support this behavior: when the visibility query is executed on a
list of primitives that are not composed into a BVH, the custom intersector
replaces the respective call to \texttt{intersect} inside the traveral loop.
When the query is however executed on a list of BVHs, the custom intersector is
passed on to the BVH intersection routine, which will replace its respective
calls to \texttt{intersect}. Discerning the two implementations is done at
compile time using the ``substitution failure is not an error'' (SFINAE)
pattern.

\section{Use Cases}
Custom intersectors can e.g.\ be used to implement the use cases from
\autoref{fig:teaser}. 3-d models often come with separate alpha masks stored in
texture images that are used to carve out details from the otherwise coarse
geometry that serves as an impostor or a billboard. This can easily be
implemented by providing a custom intersector which stores a pointer to the
texture and texture coordinate lists as member variables. The custom intersector
provides an \texttt{operator()} that intercepts interactions with the surface
geometry and performs a texture lookup using the barycentric coordinates at the
hit point to determine if the surface was actually hit. It will then manipulate
the hit point based on the alpha information from the mask texture and only then
return the hit record:
\begin{lstlisting}[
    language=C++,
    captionpos=b,
    label=lst:generic]
template <typename Ray>
auto operator()(
        Ray const& ray,
        basic_triangle<3, float> const& tri
        )
{
    auto hr = intersect(ray, tri);

    auto const& tex = textures[hr.geom_id];
    vec2 coord = lerp(
            tex_coords[hr.prim_id * 3],
            tex_coords[hr.prim_id * 3 + 1],
            tex_coords[hr.prim_id * 3 + 2],
            hr.u, hr.v
            );
    vec4 color = tex2D(tex, coord);

    hr.hit &= color.w >= .01f;

    return hr;
}
\end{lstlisting}

Another use case that is depicted in the second from right image of
\autoref{fig:teaser} is procedural alpha masking which can be implemented in a
similar way that alpha masking with texture images is implemented, but uses a
procedural to determine visibility when passed the \texttt{uv} coordinates.

A third use case that is depicted in the right image of~\autoref{fig:teaser} is
spotting and visualizing performance issues with the intersection routine or the
BVH traversal. Therefore, a custom intersector is implemented that just counts
the number of interactions. A full implementation might look like this:
\begin{lstlisting}[
    language=C++,
    captionpos=b,
    label=lst:generic]
struct bvh_costs : basic_intersector<bvh_costs> {
    using basic_intersector<
        bvh_costs
        >::operator();

    // Intercept and count ray/aabb tests
    template <typename Ray, typename ...Args>
    auto operator()(
            Ray const& ray,
            aabb const& box,
            Args&&... args
            )
    {
        ++num_boxes;
        return intersect(
                ray,
                box,
                std::forward<Args>(args)...
                );
    }

    // Intercept and count ray/triangle tests
    template <typename Ray>
    auto operator()(
            Ray const& ray,
            basic_triangle<3, float> const& tri
            )
    {
        ++num_tris;
        return intersect(ray, tri);
    }

    unsigned num_boxes = 0;
    unsigned num_tris  = 0;
};
\end{lstlisting}
After performing ray vs. BVH traversal with the \texttt{bvh\_costs} intersector,
the number of interactions with bounding boxes and with triangles will be stored
in the variables \texttt{num\_boxes} and \texttt{num\_tris} respectively and can
be used for further analysis or visualization. Note that the \texttt{aabb}
intersection function has a special interface as it also takes the inverse ray
direction as a parameter. This is opaquely handled by the custom intersector,
which just passes any additional parameters on to the intersection function
using variadic templates. The implementation provides an example of how to
support intersection functions with a non-standard interface.

\section{Conclusion}
We presented custom intersectors as a way to augment the ray tracing template
library Visionaray. Custom intersectors or similar concepts are supported by
many ray tracing libraries, but due to the static and compile time nature of
Visionaray, the feature is supported in a unique way that is different from the
usual function pointer approach that other libraries implement. We presented
the API to make use of custom intersectors, some implementation notes, and also
some sample implementations that exemplarily present how to use that feature
with Visionaray.

\footnotesize
\bibliographystyle{eg-alpha}
\bibliography{egbibsample}

\newcommand{\etalchar}[1]{$^{#1}$}
\begin{thebibliography}{\uppercase{WWB{\etalchar{*}}14}}

\bibitem[AL09]{aila:2009}
\textsc{Aila T., Laine S.}:
\newblock Understanding the efficiency of ray traversal on {GPU}s.
\newblock In \emph{Proceedings of the Conference on High Performance Graphics
  2009} (New York, NY, USA, 2009), HPG '09, ACM, pp.~145--149.

\bibitem[PBD{\etalchar{*}}10]{parker:2010}
\textsc{Parker S.~G., Bigler J., Dietrich A., Friedrich H., Hoberock J., Luebke
  D., McAllister D., McGuire M., Morley K., Robison A., Stich M.}:
\newblock {OptiX}: A general purpose ray tracing engine.
\newblock \emph{ACM Trans. Graph. 29}, 4 (July 2010), 66:1--66:13.

\bibitem[WWB{\etalchar{*}}14]{wald:2014}
\textsc{Wald I., Woop S., Benthin C., Johnson G.~S., Ernst M.}:
\newblock Embree: A kernel framework for efficient {CPU} ray tracing.
\newblock \emph{ACM Trans. Graph. 33}, 4 (July 2014), 143:1--143:8.

\bibitem[ZHL17]{zellmann:2017}
\textsc{Zellmann S., Hoevels M., Lang U.}:
\newblock Ray traced volume clipping using multi-hit {BVH} traversal.
\newblock In \emph{Proceedings of Visualization and Data Analysis (VDA)}
  (2017), IS\&T.

\bibitem[ZHL19]{zellmann:2019}
\textsc{Zellmann S., Hellmann M., Lang U.}:
\newblock A linear time {BVH} construction algorithm for sparse volumes.
\newblock In \emph{Proceedings of the 12th IEEE Pacific Visualization
  Symposium} (2019), IEEE.

\bibitem[ZL17]{zellmann:2017c}
\textsc{Zellmann S., Lang U.}:
\newblock C++ compile time polymorphism for ray tracing.
\newblock In \emph{Proceedings of the Conference on Vision, Modeling and
  Visualization} (Goslar Germany, Germany, 2017), VMV '17, Eurographics
  Association, pp.~129--136.

\bibitem[ZML19]{zellmann:2019c}
\textsc{Zellmann S., Meurer D., Lang U.}:
\newblock Hybrid grids for sparse volume rendering.
\newblock In \emph{IEEE VIS 2019 - Short Papers} (2019).

\bibitem[ZPL15]{zellmann:2015}
\textsc{Zellmann S., Percan Y., Lang U.}:
\newblock {Advanced texture filtering: a versatile framework for reconstructing
  multi-dimensional image data on heterogeneous architectures}.
\newblock In \emph{Visualization and Data Analysis 2015} (2015), Kao D.~L., Hao
  M.~C., Livingston M.~A., Wischgoll T., (Eds.), vol.~9397, International
  Society for Optics and Photonics, SPIE, pp.~110 -- 120.

\bibitem[ZSL18]{zellmann:2018}
\textsc{Zellmann S., Schulze J.~P., Lang U.}:
\newblock Rapid k-d tree construction for sparse volume data.
\newblock In \emph{Eurographics Symposium on Parallel Graphics and
  Visualization} (2018), Childs H., Cucchietti F., (Eds.), The Eurographics
  Association.

\bibitem[ZSL19]{zellmann:2019b}
\textsc{{Zellmann} S., {Schulze} J.~P., {Lang} U.}:
\newblock Binned k-d tree construction for sparse volume data on multi-core and
  {GPU} systems.
\newblock \emph{IEEE Transactions on Visualization and Computer Graphics}
  (2019), 1--1.

\bibitem[ZWL17]{zellmann:2017b}
\textsc{Zellmann S., Wickeroth D., Lang U.}:
\newblock Visionaray: A cross-platform ray tracing template library.
\newblock In \emph{Proceedings of the 10th Workshop on Software Engineering and
  Architectures for Realtime Interactive Systems (IEEE SEARIS 2017)} (in press,
  2017), IEEE.

\end{thebibliography}

\end{document}